# A simulator for maxillo-facial surgery integrating 3D cephalometry and orthodontia

Bettega G [1], Payan Y [2], Mollard B [2], Boyer A [3], Raphaël B [1] and Lavallée S [3]

1 : Service de Chirurgie Plastique et Maxillo-Faciale - Centre Hospitalier Universitaire de Grenoble - France

2 : Laboratoire TIMC / IMAG - Université Joseph Fourier de Grenoble - France

3 : PRAXIM - La Tronche - France

*Corresponding author* : Yohan Payan

*Address* : Laboratoire TIMC, équipe GMCAO - Institut Albert Bonniot, Faculté de Médecine, 38706 La Tronche Cedex, FRANCE.

*Phone*   : +33 4-76-54-95-22

*Fax*      : +33 4-76-54-95-55

*E-mail*  : Yohan.Payan@imag.fr

# Abstract

*Objectives :*

This paper presents a new simulator for maxillo-facial surgery, that gathers the dental and the maxillo-facial analyses together into a single computer-assisted procedure. The idea is first to propose a repositioning of the maxilla, via the introduction of a 3D cephalometry, applied to a 3D virtual model of the patient's skull. Then, orthodontic data are integrated into this model, thanks to optical measurements of teeth plaster casts.

*Materials and Methods :*

The feasibility of the maxillo-facial demonstrator was first evaluated on a dry skull. To simulate malformations (and thus to simulate a "real" patient), the skull was modified and manually cut by the surgeon, in order to generate a given maxillo-facial malformation (with asymmetries in the sagittal, frontal and axial planes).

*Results :*

The validation of our simulator consisted in evaluating its ability to propose a bone repositioning diagnosis that will put the skull as it was in its original configuration. A first qualitative validation is provided in this paper, with a 1.5-mm error in the repositioning diagnosis.

*Conclusions :*

These results mainly validate the concept of a maxillo-facial numerical simulator that integrates 3D cephalometry and guarantees a correct dental occlusion.

# INTRODUCTION

Planning cranio-facial surgical procedures, particularly orthognathic surgery, requires the integration of multiple and complex data gathered from different sources: clinical examination (anthropometry), orthodontic (dental models), radiological (cephalometry) and intra-operative data (constraints and position information). This heterogeneity makes the therapeutic decision difficult, particularly in asymmetrical dysmorphosys. This is the reason why several three-dimensional (3D) surgical analysis, simulation software and methods have been developed[3,7,9,10,11,13,14,15,17,18,19,20]. The pioneers in this field were Marsh and Vannier[13,18,19]. According to Cutting [7], a surgical simulation program has to be built with three functions. First, one must be able to cut a model of the skull respecting surgical procedures. Second, mobilizing the bone segments with six degrees of freedom is required. The third function is the creation of a 3D cephalometric analysis. To these functions, one must associate the necessity of adapting to the limitations imposed by anatomical or physiological characteristics of the area (preserving vessels, nerves, etc.). It is also important to integrate a soft tissue simulation in this bone analysis.

Another important challenge is transferring these 3D data in the operating room, in order to simplify the surgical procedure with computer [4,5]. Even if the technology is rapidly improving, the simulations proposed are still rudimentary [17], and do not respect previous 3D cephalometric analyses. 3D cephalometric analysis is a real problem and very few publications are available. Major difficulties are the high volume of data to be processed by the computer and the lack of cranio-facial normative 3D data. Altobelli [1] has discussed the use of anthropometric data or the extrapolation of two-dimensional data. Marsh [13] considers that this extrapolation is adequate in symmetrical dysmorphosis but cannot be applied to cranio-facial problems or to asymmetrical abnormalities.

Surgical simulation is usually performed in two environments: digital graphics workstations and solid life-size skull facsimiles. Digital graphics workstations allow multiple simulated operations without degradation of the database, and theoretically allow combination of osseous and soft-tissue simulation. The digital data format facilitates quantitative analysis of simulation and outcome, but it is very difficult to define the dental occlusion with adequate accuracy. From this point of view, stereolithographic models are more concrete for the surgeon [6]; the occlusal problem can be solved, and implants can be prepared prior to surgery. However, the manipulation is destructive, and the cost and fabrication time of the model are shortcomings of this procedure.

This paper deals with a 3D cephalometric analysis system and a surgical simulator for orthognathic surgery that integrates the advantages of both environments (graphic and facsimiles). The simulator is based on the integration of dental models and 3D cephalometry.

## OBJECTIVES

### State of the art

Orthognathic surgery deals with face dysmorphosis arising from congenital malformations or accidents [16]. For example, in the case of mandibular prognathism (dento-facial deformity of the lower third of the face resulting from excess mandibular growth), orthognathic surgical treatment is required to correct the occlusion (dental position) with an osteotomy of the mandible [2]. For this treatment, and for many other orthognathic surgical acts, surgeons usually start from (1) teeth plaster casts and (2) sagittal, frontal and/or axial 2D radiographies of the patient's head.

Teeth plaster casts are used to plan the osteotomy phase: both casts (mandible and maxilla) are manually cut to simulate (1) a correct positioning of the maxilla in relation to some specific face anatomical landmarks (by the mean of a facial bow study), and (2) a correct positioning of the mandible in relation to the maxilla, to guarantee a normal dental occlusion. During this cutting procedure, resin splints (called *intercuspidation splints*) are built from plaster casts, gathering dental occlusion prints for initial (actual maxilla and mandible), intermediary (actual mandible and maxilla cut) and final (maxilla and mandible cut) plaster casts positions. Those splints are used during surgery, as references for maxillary and mandibular osteotomies.

In parallel to this dental planning, surgeons can make a simplified 2D Delaire cephalometry [8], *i.e.* compute, from sagittal and/or facial radiographic tracings, the desired displacements of the maxilla and the mandible. This is achieved first by placing specific anatomical landmarks onto the radiography. Then, thanks to tracing papers, suited displacements of mandibular and maxillary landmarks, in relation to the rest of the skull, are manually measured (figure 1).

This cephalometric diagnosis is then compared to the displacements provided by the orthodontic facial bow study.

INSERT FIGURE 1 AROUND HERE

As can be drawn here, two parallel procedures are required for the planning of orthognathic surgery: dental analysis and maxillo-facial analysis, both working on different kind of supports. Moreover, the decision phase only occurs at the end of each procedure, which means a waste of time if both planning are not compatible and if procedures have to be driven again.

The aim of the work presented in this paper is to gather those two surgical planning into a single computer-assisted procedure, that integrates information from surgeons, about repositioning of the bone structures, and information from orthodontists, about optimal dental occlusion. The idea is first to use orthopedic knowledge to propose a repositioning of the maxilla, via the introduction of a 3D cephalometry, applied on a 3D virtual model of the patient skull. Then, orthodontic data are integrated into this model, thanks to measurements of teeth plaster casts, with the use of a 3D localizing system. For this, only the final suited position of the mandible, in relation to the maxilla, is taken into account. No cast-cutting phase is thus required, which makes the procedure lighter.

## MATERIALS AND METHODS

### Choice for the patient

The feasibility of our demonstrator was first evaluated on a dry skull. This skull was a "standard" one without any noticeable maxillo-facial dysmorphosys (figure 2, left panel). This choice was motivated by our wish to be able to quantify the repositioning diagnosis proposed by the simulator, which necessitates having a knowledge of the "normal" maxillary and mandibular positions. To simulate malformations (and thus to simulate a "real" patient), the skull was modified and manually cut by the surgeon (figure 2, right panel), in order to generate a given maxillo-facial dysmorphosys (with asymmetries in the sagittal, frontal and axial planes). Before this cutting phase, two parallel tubes were fixed between the forehead and the mandible.

INSERT FIGURE 2 AROUND HERE

As can be seen on figure 2 (right panel), each tube had to be cut into three parts, in order to allow the manual bone structure-cutting phase. The validation of our simulator will thus consist in evaluating its ability to propose a repositioning diagnosis that re-aligns each parts of the two tubes, as in the original skull configuration.

**Data acquisition and 3D reconstruction of the patient's skull**

Horizontal Computer Tomography (CT) slices were collected for the whole skull (helical scan with a 3mm pitch and slices reconstructed each 1.5 mm). The *Marching Cubes* algorithm [12] has been implemented to reconstruct the skull from CT slices. Before running this reconstruction process, tools (erasers) can be used to clean specific slices, and a threshold value for the reconstructed isosurface has to be chosen (figure 3 – top panel – shows a snapshot of the PC platform software). Then, the process automatically builds the virtual 3D model (figure 3, low panel).

INSERT FIGURE 3 AROUND HERE

**3D cephalometry**

The third dimension brings to cephalometric analysis the advantage of taking into account the data given by frontal, sagittal and axial studies in a single step. It allows the integration of the problems of facial asymmetry and occlusal plane horizontality into the profile analysis. Apart from the implementation, the main problem in 3D cephalometry is the standardization and the reference to the norms that exist in bidimensional cephalometries. Instead of creating a new 3D analysis, the idea was to transpose the data of 2D cephalometry in the third dimension. Our approach consists in a 3D extrapolation of the simplified Delaire analysis and is illustrated in Figure 4.

INSERT FIGURE 4 AROUND HERE

This analysis is adapted to the third dimension so that the reference of standards existing in the sagittal plane is respected. The norms in the other dimensions are theoretically easy to define. It is a matter of respecting the horizontality in the frontal plane and the symmetry in relation with the sagittal median plane.

The surgeon is therefore asked to manually position each of the points listed in figure 4 onto the virtual model of the patient's skull (figure 5 top panels). Starting from those cephalometric points, an automatic analysis procedure provides specific lines and specific planes (figure 5 low panel), which will be used for the determination of a repositioning diagnosis.

INSERT FIGURE 5 AROUND HERE

Before this diagnosis, pixels of the 3D model belonging to the maxilla / mandible block have to be labeled, as the repositioning diagnosis will be applied to these points. For this, a virtual osteotomy is manually simulated, which separates the skull model into two groups of points (figure 6).

INSERT FIGURE 6 AROUND HERE

As shown in Figure 6, the virtual osteotomy is performed using a parallelepiped-cutting pattern, that is interactively placed on the skull model and dimensioned with the manipulation tools provided by the software. These tools are sufficient to obtain a realistic model of surgical cutting off.

## RESULTS

**Maxilla repositioning diagnosis**

Maxilla repositioning is totally driven by the cephalometric analysis, according to three following constraints (see figure 4 for the name of planes and points):

- NP point is moved to fit the theoretical NP point position, computed from cephalometric analysis (onto the intersection between $CF_1$ plane and the sagittal median (SM) plane).
- $CF_7$ plane is moved to fit the theoretical $CF_7$ plane.
- A given point chosen at the intersection between the two patient incisors is moved to be projected onto the sagittal median plane.

Following those constraints, a global displacement of the maxillary structure is computed in terms of translation and rotation. Figure 7 (low panel) plots the corresponding repositioning diagnosis (*N.B.* : mandible position, in relation to the maxilla, remains constant during this operation).

INSERT FIGURE 7 AROUND HERE

**Mandible repositioning diagnosis**

As the mandible repositioning has to integrate dental occlusion constraints, it was decided to let it be totally driven by dental diagnosis. Cephalometric points and planes resulting from our analysis were here only used to determine the real occlusion plane ($CF_7$), between maxilla and mandible (in order to label points with their belonging to the maxilla or to the mandible).

Like in usual treatments, orthodontic diagnosis was carried out on teeth plaster casts. The only concern was the position of the mandible in relation to the maxilla, in terms of optimal dental occlusion. Contrary to standard orthognathic procedures, teeth plaster casts had not to be cut, as no concern was needed in terms of maxilla positioning in reference to the rest of the skull.

The initial and final splints were used for the integration of the orthodontic diagnosis into the virtual model of the patient's skull (the intermediary splint is thus removed from the procedure). Inserted into the teeth plaster casts, these splints respectively replicate the current and desired positions of the mandible, in relation to the maxilla. A three-dimensional optical localizer (Optotrak$^{TM}$, NorthernDigital) is therefore used to quantitatively measure the corresponding displacement, from current (figure 8, left panel) to desired position (figure 8, right panel).

INSERT FIGURE 8 AROUND HERE

This measurement consists in a global transformation matrix, corresponding to the desired translation and rotation that has to be applied to the mandible. This transformation being expressed in the localizer referential, it has to be transferred into the CT scans referential, *i.e.* into the virtual model space. To do this, an object - visible in both modalities (localizer space *versus* CT space) - has been introduced into the procedure. This object is a pair of aluminum tubes, fixed onto the initial intercuspidation splint (figure 9).

INSERT FIGURE 9 AROUND HERE

Those tubes being aluminum made, they can be detected on CT scans, if they are inserted inside the patient's mouth during the CT recordings (figure 10, right panel). Moreover, they can also be detected and located into the optical localizer referential (figure 10, left panel). Therefore, a simple matching algorithm enables us to compute the transformation from one referential to the other, and thus transpose the mandible correction into the virtual model space (figure 11, low panel).

INSERT FIGURE 10 AROUND HERE

INSERT FIGURE 11 AROUND HERE

## DISCUSSION

The repositioning diagnosis simulation presented on figure 11 validates the feasibility of our simulator, as each part of the two tubes is qualitatively re-aligned with the other ones.

These tubes show a maximal deviation of 2 degrees between their axes, which roughly corresponds to a 1.5-mm error in the repositioning procedure. These results mainly validate the concept of a maxillo-facial numerical simulator that (1) integrates 3D cephalometry, (2) guarantees a correct dental occlusion, and (3) proposes a semi-automatic diagnosis for maxilla and mandible repositioning. However, the obtained errors are not completely satisfying as the aim for orthognathic surgery would be to have a precision below one millimeter. The next step of this work would thus consist in a quantitative evaluation of each point of the global procedure: precision of the computer-assisted 3D cephalometry, precision of the dental occlusion measurement, repeatability of the process for the same patient and/or for different kind of pathologies. Moreover, a comparison of the simulator diagnosis with osteotomies provided by surgeons on classical procedures, without computer-assisted techniques, would be another important evaluation factor.

## CONCLUSION

In this paper we presented a first evaluation of the feasibility of computer-assisted techniques for maxillo-facial surgery, integrating 3D cephalometry and orthodontic information. First tests have been carried out on a dry skull. This skull was manually cut, to simulate a pathological patient. The repositioning diagnosis proposed by the maxillo-facial simulator was evaluated through re-alignment of tubes that were fixed on the cadaver skull before the simulation of the patient dysmorphosys. A first clue in the efficiency of the simulator was provided and discussed. Clinical tests have now to be carried out on real patients, and with different types of pathologies. Then, the next phase of this work will consist in transferring the simulated repositioning diagnosis into the operating room.

# ACKNOWLEDGEMENTS

This work was financially supported by the european project IGOS HC1026HC.

**FIGURE CAPTIONS**

Figure 1: Delaire cephalometry on sagittal radiographic tracings: computation of maxilla and mandible repositioning thanks to tracing papers.

Figure 2: A "standard" dry skull (left) manually cut to simulate malformations (right).

Figure 3: Software interface: horizontal CT slices (top) and the corresponding reconstructed 3D model of the patient's skull (down).

Figure 4: 3D extrapolation (right) of the simplified Delaire analysis (left).

Figure 5: Positioning of the anatomical landmarks (top panels) and the corresponding 3D cephalometric analysis (low panels).

Figure 6: A cube to simulate a virtual osteotomy that separates the maxilla/mandible block from the rest of the skull.

Figure 7: A diagnosis for the repositioning of the maxilla/mandible block (low panels), automatically obtained from the 3D cephalometric analysis.

Figure 8: Intercuspidation splints inserted into teeth plaster casts, localized by the means of optical rigid bodies. Left: actual position; right: desired dental occlusion.

Figure 9: Initial intercuspidation splint with two aluminum tubes.

Figure 10: Measurements of the aluminum tubes axes in the localizer referential (left) and in the CT scans referential (right).

Figure 11: Integration of the orthodontic mandible repositioning diagnosis (low panels) into the virtual model of the patient's skull.

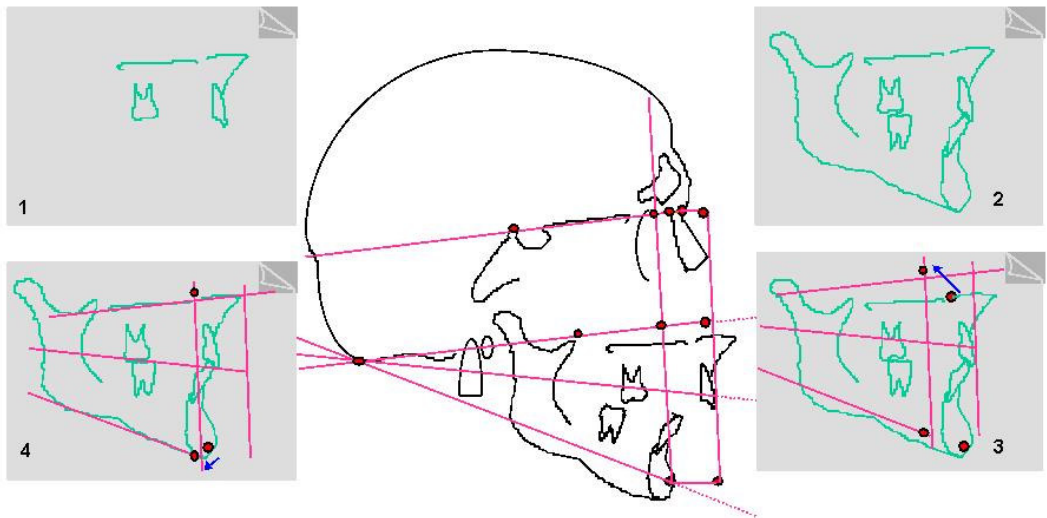

Figure 1

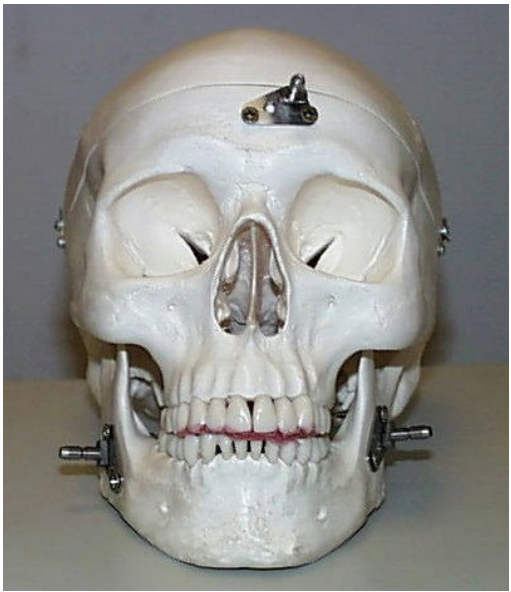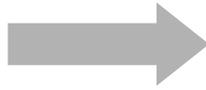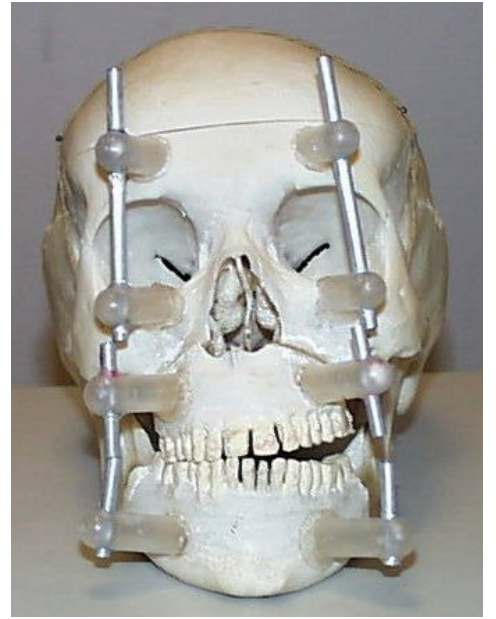

Figure2

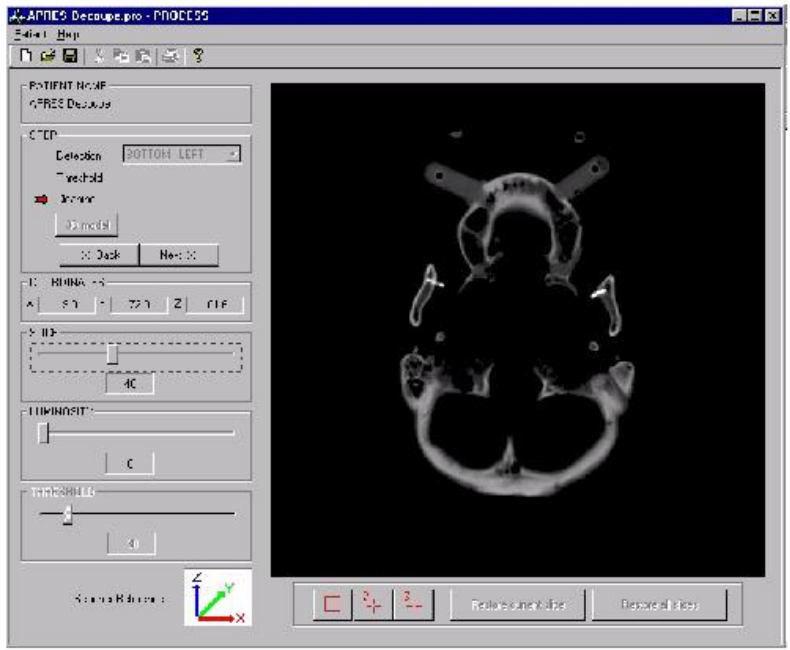
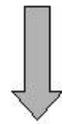
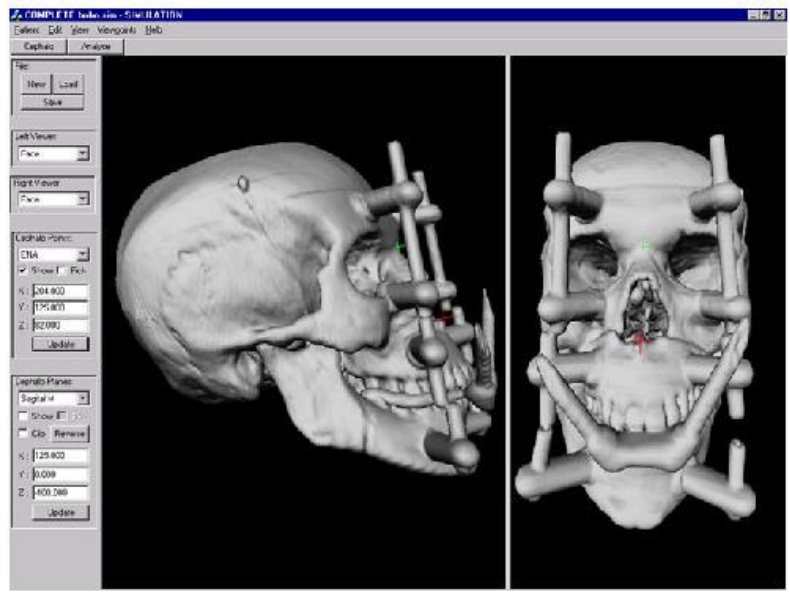

Figure 3

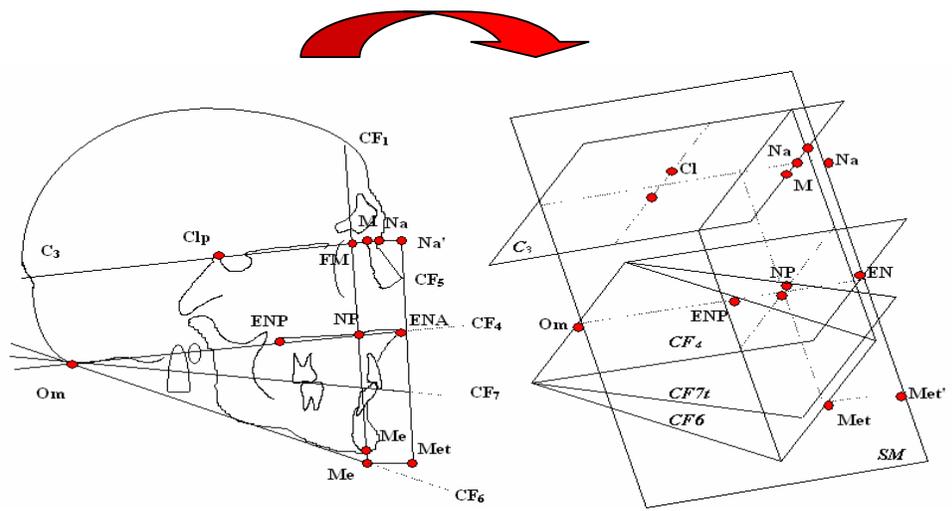

Figure 4

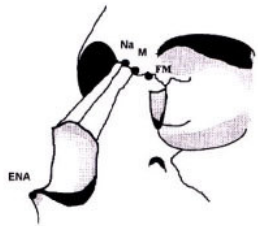 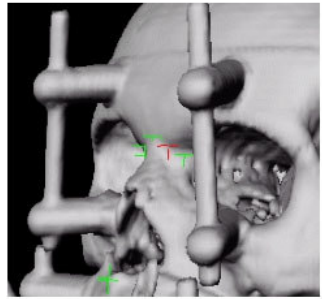

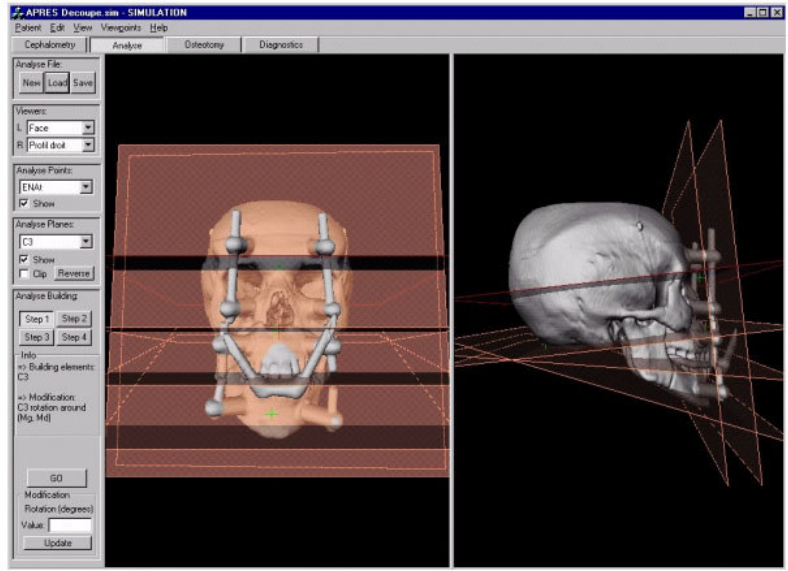

Figure 5

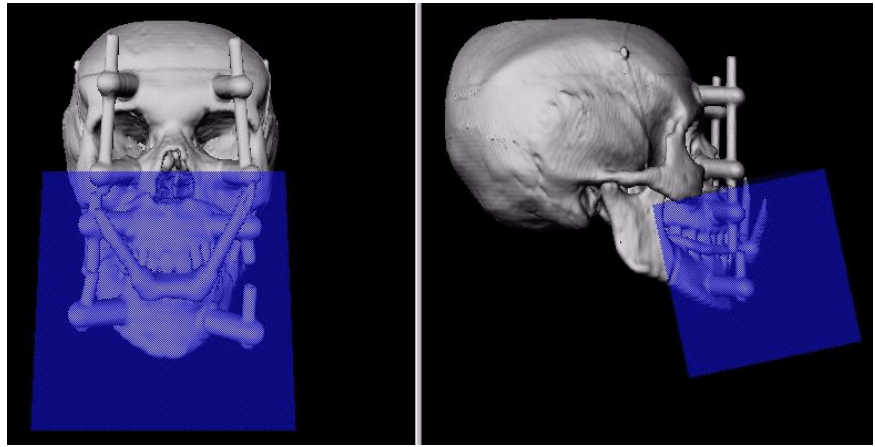

figure 6

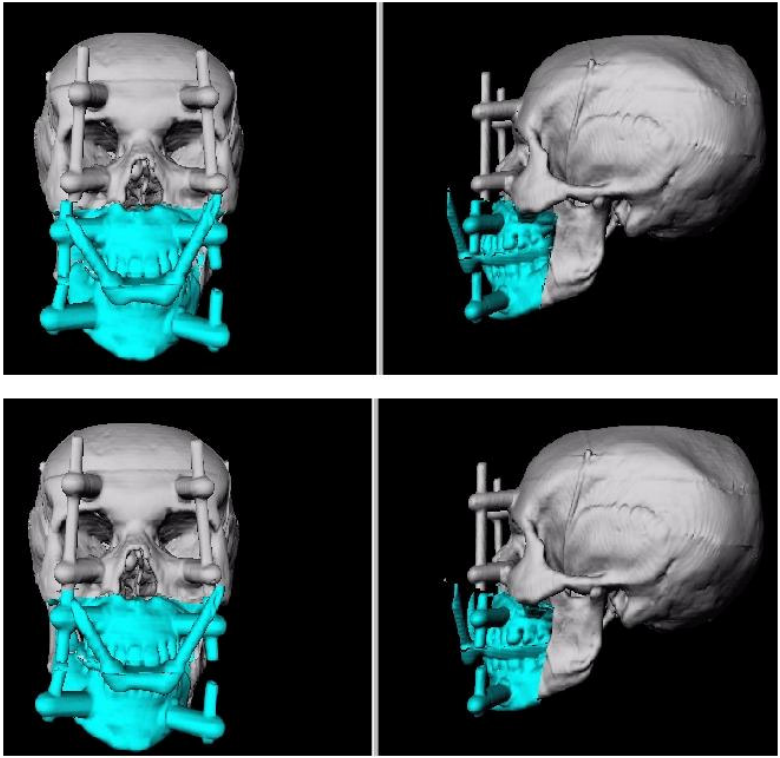

figure 7

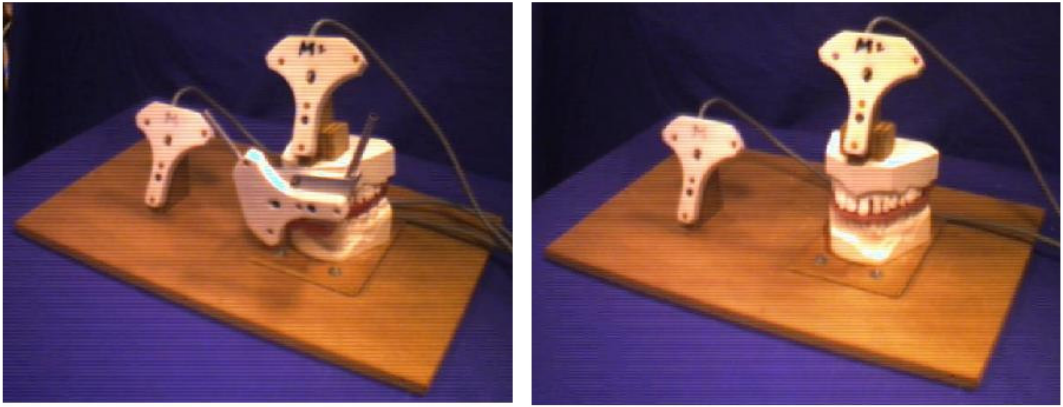

figure 8

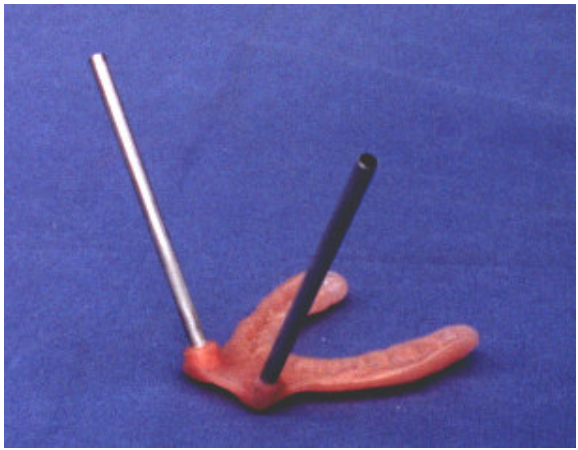

figure 9

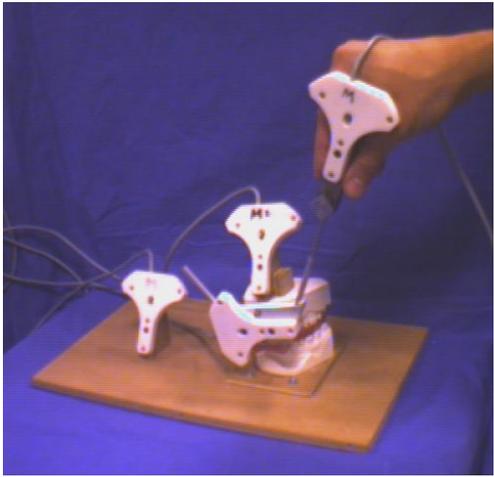 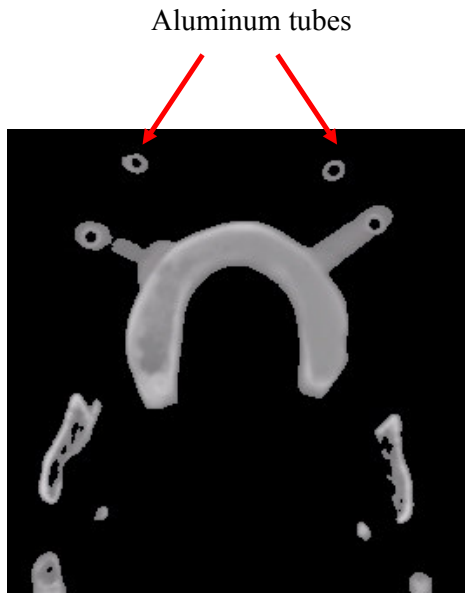

Figures 10

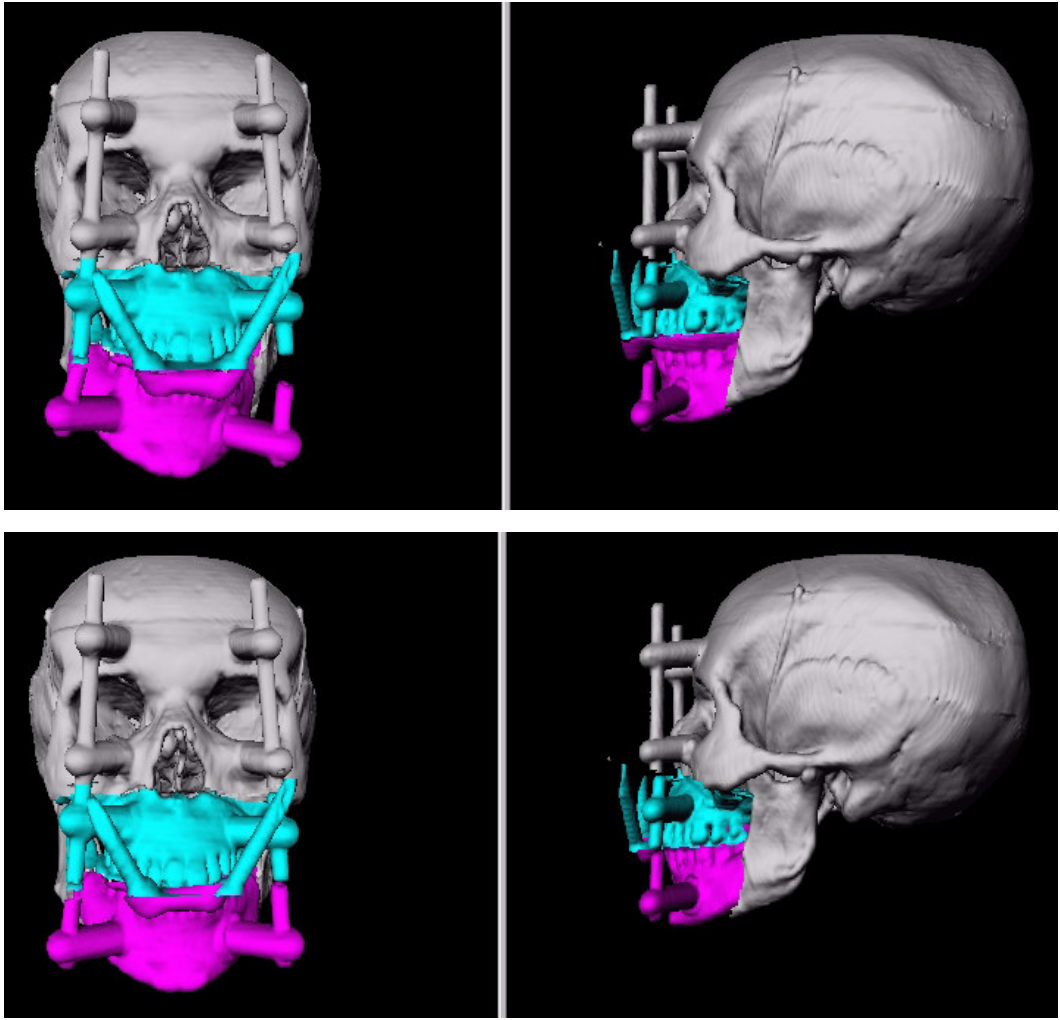

figure11